\documentclass[10pt,letterpaper,twocolumn]{article} 
\usepackage{graphicx}
\usepackage{pstricks}

\usepackage{ol2}
\usepackage[draft]{hyperref}
\usepackage{amsmath}

\begin{document}

\twocolumn[ 

\title{Strong saturation absorption imaging \\ of dense clouds of ultracold atoms}


\author{G. Reinaudi,$^{1}$ T. Lahaye,$^{1,2}$ Z. Wang, $^{1,3}$ and D. Gu\'ery-Odelin$^{1,*}$}

\address{
$^1$Laboratoire Kastler Brossel, Ecole Normale Sup\'erieure,  24 rue
Lhomond,  F-75231 Paris Cedex 05, France
\\
$^2$5. Physikalisches Institut, Universit\"at Stuttgart,
Pfaffenwaldring 57, D-70550 Stuttgart, Germany
\\
$^3$Institute of Optics, Department of Physics, Zhejiang University,
Hangzhou 310027, China
\\
$^*$Corresponding author: dgo@lkb.ens.fr}

\begin{abstract}

We report on a far above saturation absorption imaging technique
to investigate the characteristics of dense packets of ultracold
atoms. The transparency of the cloud is controlled by the incident
light intensity as a result of the non-linear response of the
atoms to the probe beam. We detail our experimental procedure to
calibrate the imaging system for reliable quantitative
measurements, and demonstrate the use of this technique to extract
the profile and its spatial extent of an optically thick atomic
cloud.
\end{abstract}


 ] 

\noindent

Recently there has been a resurgent interest in the production of
dense samples~\cite{cornellpmo,Townsend} containing a large number
of cold neutral atoms with highly compressed magneto-optical traps
(MOT) ~\cite{weiss,prentiss}. These studies, combined with optical
trapping, opened the way to a simplified and very rapid production
of Bose-Einstein condensates~\cite{weiss2}, and may also play a
key role in the production of a cw atom laser based on the
periodical coupling of atomic packets into a magnetic guide,
yielding a promising starting point for evaporative
cooling~\cite{prl04}.

For dense clouds, an important issue is the reliability of the
method used to extract the atomic densities. The predominant
imaging techniques for dilute samples, low-intensity fluorescent
and absorption imaging, turn out to be unreliable when probing a
dense atomic packet~\cite{or1,him}. The former critically depends
on the value of the illuminating intensity, the frequency and the
repartition of atoms between the different Zeeman sub-levels. The
latter poses a problem as soon as the optical depth, proportional
to the column density of atoms along the probe direction, is on
the order of 3 to 4, because of both (i) the electronic noise of
each pixel of the charge-coupled-device (CCD) camera and (ii) the
digitalization of the signal of the weak amount of light that
remains after the propagation through the cloud.

An optically thick cloud can also be probed with off-resonant
light or by implementing the phase contrast imaging
technique~\cite{or1}. The reduction of the scattering cross
section that results from the non-resonant probe is favorable.
However, the sample becomes dispersive and behaves like a
gradient-index lens. Such a technique is fruitful for qualitative
or differential measurements, but is difficult to use for
quantitative purposes.

To circumvent those drawbacks the group of D. Weiss has
successfully implemented a fluorescence imaging technique very far
above saturation intensity \cite{weiss}: the probe intensity was
larger by more than three orders of magnitude than the saturation
intensity. In this regime, all atoms, independently of their
Zeeman sub-level distribution, spend half of the time in the
excited state. This method allows to probe a cloud with a very
large optical depth, but requires a dramatically high incident
intensity and a subtle alignment of the two counter propagating
beam used to drive the fluorescence.

 In this article, we
report on our realization of a robust, accurate and reliable far
above saturation intensity absorption imaging aimed at
investigating such dense atomic samples. Its implementation is
straightforward and, by contrast with the far above saturation
fluorescence imaging technique, the probing does not required the
use of a powerful laser.

Indeed, in our experiment, the probe light is provided only by
diode lasers. A semiconductor slave laser is injection-locked to a
0.5 MHz linewidth Distributed Bragg Reflector (DBR) master diode
laser, and spatially filtered by a pinhole. The purpose of this
arrangement is to benefit from a narrow linewidth probe tuned on
the transition $^{87}{\rm Rb}, 5^2S_{1/2}\rightarrow 5^2P_{3/2}$
while having a relatively large power (30 mW) available to probe
the atoms. An acousto-optic modulator placed before the pinhole is
used to produce light pulses as short as 250 ns. The shadow cast
by the atoms on the resonant probe beam is imaged on a CCD camera,
with an optical resolution of 7 $\mu$m.

 The response of the atoms,
i.e. the population driven in the excited state by the imaging
laser beam, depends on the {\em effective} saturation intensity
$I^{\rm sat}_{\rm eff}=\alpha^* I^{\rm sat}_{\rm 0}$, where
$I^{\rm sat}_{\rm 0}$ is the saturation intensity for the
corresponding two-level transition ($I^{\rm sat}_{\rm 0}=1.67$
mW/cm$^2$ for rubidium 87). The dimensionless parameter $\alpha^*$
accounts for corrections due to the specific conditions in which
images are taken: the polarization of the imaging beam, the
structure of the excited state and the different Zeeman sub-level
populations of the degenerate ground state of the optical
transition.

In order to extract the spatial atomic density $n(x,y,z)$ of the
cloud, we acquire as usual three images: $I_{\rm w}(x,y)$ with the
atoms and probe beam on, $I_{\rm wo}(x,y)$ without the atoms and
probe on, and $I_{\rm dark}(x,y)$ without atoms and probe off.
From those images, we work out for each pixel $(x,y)$, the light
intensity $I_{\rm f}(x,y)=I_{\rm w}(x,y)-I_{\rm dark}(x,y)$ (resp.
$I_{\rm i}(x,y)=I_{\rm wo}(x,y)-I_{\rm dark}(x,y)$) of the imaging
beam in presence (resp. absence) of atoms by removing the
contribution of the background light illumination taken in the
absence of the detection beam.

The Beer's law in presence of saturation effect and for a resonant
incident light can be recast in the form:
\begin{equation}
\frac{dI}{dz}=-n\frac{\sigma_0}{\alpha^*}\frac{1}{1+I/I^{\rm
sat}_{\rm eff}}I\equiv-n\sigma(I)I \label{Equation_Beer}
\end{equation}
where $\sigma_0 = 3 \lambda^2 / 2 \pi$ is the resonant
cross-section for a two-level atom, and $\sigma(I)$ the effective
cross section including saturation correction. From Eq.
(\ref{Equation_Beer}), one readily obtains the expression for the
optical depth:
\begin{equation}
od_{\rm 0}(x,y)  \equiv  \sigma_0\int \; n(x,y,z) \; dz =
f(x,y;\alpha^*). \label{Equation_Odepth}
\end{equation}
where $f(x,y;\alpha^*)$ is defined by:
\begin{equation}
f(x,y;\alpha^*) =-\alpha^* \ln \left( \frac{I_{\rm f}(x,y)}{I_{\rm
i}(x,y)}\right) + \frac{I_{\rm i}(x,y) - I_{\rm f}(x,y)}{I^{\rm
sat}_{\rm 0}}, \label{Equation_f}
\end{equation}
The optical density is defined by: $\delta_{\rm 0}(x,y) = - \ln
\left(I_{\rm f}(x,y)/I_{\rm i}(x,y)\right)$. We have deliberately
chosen a definition of the optical depth $od_{\rm 0}(x,y)$ that
does not include the parameter $\alpha^*$ so that it depends only
on the characteristics of the atomic cloud. However, to extract
this quantity from a set of images, one needs to know the
parameter $\alpha^*$.

For low-intensity absorption imaging ($I_{\rm i}(x,y)\ll I^{\rm
sat}_{\rm 0}$), the optical depth involves only the ratio of the
intensities $I_{\rm f}$ and $I_{\rm i}$: $od_0(x,y) \simeq
\alpha^*\delta_{\rm 0}(x,y)$. The unknown parameter $\alpha^*$
still needs to be determined independently. The strong saturation
imaging technique takes advantage of the reduction of the
effective cross-section $\sigma(I)$ when $I \gg \alpha^*I^{\rm
sat}_{\rm 0}$. In this limit, the optical depth depends both on
$\alpha^*$ and on the value of the incident intensity $I_{\rm
i}(x,y)$, because of the non-linear atomic response (see
Eq.~(\ref{Equation_f})).

The absolute calibration is a crucial but delicate task for
fluorescence or low absorption imaging. In the context of our far
above saturation absorption imaging technique, it just consists in
determining the parameter $\alpha^*$. We proceed in the following
manner. The sample of cold atoms is generated by a compressed
elongated two-dimensional MOT. The cloud is imaged after a not too
short time-of-flight so that its maximum optical density is not
too high ($\sim 2$), which guarantees also the validity of the
low-intensity absorption imaging. As the imaging technique is
destructive, we acquire several set of images (typically five) of
a cloud always prepared in the same conditions for different
incident intensities. In practice, we vary the intensity of the
imaging beam by more than two orders of magnitude while keeping
the number of photons per pulse constant: the duration of the
pulse was varied from 250 ns ($I_{\rm i} \simeq 23$ mW/cm$^2$) to
100 $\mu$s ($I_{\rm i} \simeq 0.06$ mW/cm$^2$). Keeping the number
of absorbed photons small ($\sim$ 5 photons per atom on average)
avoids pushing and heating the cloud, thereby changing its
characteristics.

\begin{figure}[h!]

\centerline{\includegraphics[width=7cm 
]{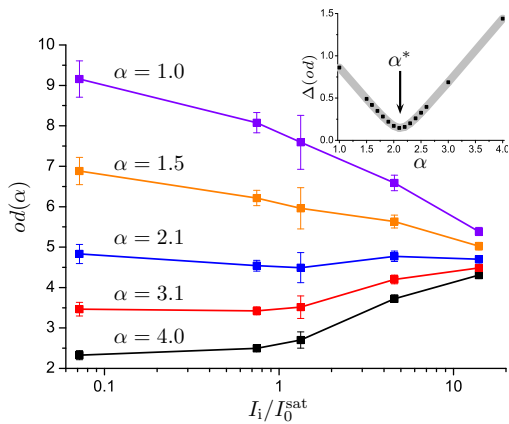}} \caption{(color online) A cloud is
imaged using different probe intensities (from $I^{\rm sat}_{\rm
0} / 15$ to $15 I^{\rm sat}_{\rm 0}$). For each image, the maximum
optical depth of the cloud $od(\alpha)$, deduced from a gaussian
fit, is calculated with several values of the unknown parameter
$\alpha$ using the function $f$. The plot represents $od(\alpha)$
as a function of the incoming intensity. The standard deviation
$\Delta (od)$ of each set of data points (see inset) exhibits a
clear minimum as a function of the parameter $\alpha$. The minimum
of
$\Delta (od)$ gives $\alpha^* = 2.12\pm 0.1$.}%
\label{Figure_Methode}
\end{figure}

In order to infer the value of the  dimensionless parameter
$\alpha^*$, we calculate, for different values of the parameter
$\alpha$ ranging from 1 to 4,  the function $f(x,y;\alpha)$ for
the set of images. We extract the amplitude $od(\alpha)$ of those
calculated optical depth using a gaussian fit (see Fig.
\ref{Figure_Methode}). There is only one value $\alpha^*$ of
$\alpha$ for which all the calculated $od(\alpha)$ are equal over
the whole range of incident intensities. Indeed, $od_0$ does not
depend on the incident probe intensity (see Eq.
\ref{Equation_Odepth}).

In practice, we infer by a least square method the value
$\alpha^*$ for which $od(\alpha)$ has a minimum standard deviation
over the whole range (more than two orders of magnitude) of
incident intensities used to image the cloud (see Fig.
\ref{Figure_Methode} inset). We find $\alpha^* = 2.12\pm 0.1$ and
a maximum optical depth $od_{\rm 0} = 4.8$ that corresponds to an
optical density of $\delta_{\rm 0} = 2.25$ as deduced from the low
intensity absorption imaging. We stress that this calibration
allows for an absolute determination of the number of atoms, its
accuracy being ultimately determined by the calibration of the
incident intensity using the CCD array.


This value for $\alpha^*$ is to be compared with the result of the
Bloch equations for the corresponding multiple level system. For
our data, the atoms are initially in the
$|g\rangle=5^2S_{1/2},F=2$ hyperfine state (5-fold degenerated)
and the $\pi$-polarized probe is resonant with the
$|e\rangle=5^2P_{3/2},F'=3$ hyperfine excited state (7-fold
degenerated). The probability to excite the $5^2P_{3/2},F'=2$ is
negligible (below 1 \%). The transition $|g\rangle \to|e\rangle$
can be considered in this limit as closed. From numerical
integration of the Bloch equations, we find out that (i) the
steady state solution is approximately valid even for the shortest
pulses that we use, and (ii) the correction factor $\alpha^*$ to
the two-level saturation intensity lies in between one and two
depending on the polarization of the probe. Experimentally, the
polarization of the beam cannot be perfectly under control because
of the slight birefringence of the viewports. In addition, a
residual magnetic field may also influence the effective value of
$\alpha^*$.

\begin{figure}[h!]
\centerline{
\includegraphics[width=6.5cm 
]{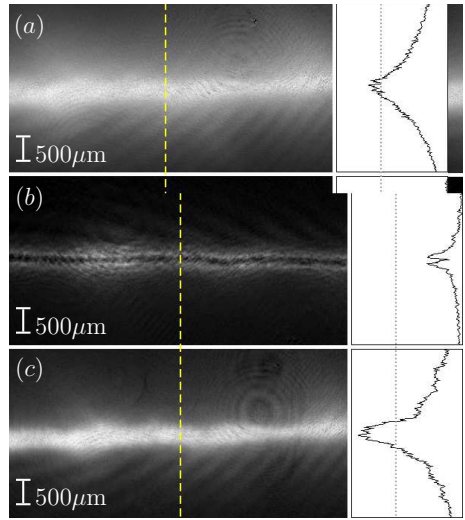} } \caption{Three images with a transverse cut
of a dense elongated cloud prepared in the same conditions. The
transverse profile is given by the dimension less function
$f(x,y;\alpha^*)/\alpha^*$ and the dashed line corresponds to the
value 3. (a) Low intensity absorption imaging. Almost all the
light is absorbed in the center of the cloud. (b) Off-resonance
absorption imaging with a probe laser blue-detuned by $1.2 \Gamma$
where $\Gamma$ is the natural linewidth of the excited state. One
observes a double peak structure resulting from a gradient-index
lens effect. (c) High intensity absorption imaging. Only the third
method shows that the profile has a double structure, with an
r.m.s size for the central peak of 300 $\mu$m, corresponding to a
peak atomic density $\sim 4(\pm 1) \times 10^{10}$ atoms/cm$^3$
and
max$(od_0)\sim$ 9.}%
\label{Figure_3Photos}
\end{figure}

After calibration, the high-intensity imaging technique is applied
to the same cloud but without time-of-flight. The transparency of
the atomic cloud is controlled by the probe intensity because of
the non-linear response of the atoms. Our numerical studies based
on the Eqs.~(\ref{Equation_Odepth}) and (\ref{Equation_f}) show
that the true profile of the cloud can be reliably inferred from
the strong saturation absorption images as soon as the incident
intensity of the probe beam is on the order of $I_{\rm i}\sim {\rm
max}(od_{\rm 0})\,I^{\rm sat}_{\rm eff}$. Note that this intensity
is much less than the one needed for a reliable high intensity
fluorescence imaging technique~\cite{weiss}. We have checked this
prediction by imaging with different incident intensities a
compressed two-dimensional MOT prepared in the same conditions. To
compress the MOT, we proceed in the following manner: the repumper
intensity is divided by 15 in 1 ms, the detuning is ramped
linearly from $-3\Gamma$ to $-9\Gamma$ in 15 ms, and the gradient
is increased from 5~G/cm to 20~G/cm in 15 ms. From our analysis
based on the strong saturation absorption, we systematically find
out a double structure (see Fig. \ref{Figure_3Photos}.c) with a
central dense region having an maximum peak atomic density on the
order of $2(\pm 1) \times 10^{11}$ atoms/cm$^3$ corresponding to
max$(od_0) \sim$ 45. Half of the atoms ($1.5\times 10^8$) remains
in the wing. The profile of those wings and their number of atoms
inferred from the low and high intensity absorption imaging
techniques perfectly coincides.

In addition, this technique is particularly well suited for the
estimation of the spatial extent of dense atomic packets with one
or more sizes very small ($\Delta x_0\leq \; 30$ $\mu$m). If $T$
is the temperature of the cloud, the size of the packet reflects
the velocity distribution after a time-of-flight duration on the
order of $\tau = (m\Delta x_0^2 / k_B T)^{1 / 2}$. For a cloud of
initial size 10 $\mu$m and a temperature of 100 $\mu$K, $\tau \sim
100$ $\mu$s. By contrast with low-intensity absorption imaging,
for which the pulse duration is on the order of the time $\tau$,
the very short pulse used for high-intensity imaging prevents
heating and permits to extract the correct profile.

We have demonstrated a high-intensity absorption imaging technique
well-suited for dense, small-size atomic clouds. This technique is
robust against small frequency and intensity variations of the
probe, and does not require an extremely large probe intensity. We
have shown how it can be reliably calibrated exploiting the
non-linear atomic response.

We thank J. Dalibard, T. Kawalec and A. Couvert for careful
reading of the manuscript. Support for this research came from the
D\'el\'egation G\'en\'erale pour l'Armement (DGA) and the Institut
Francilien de Recherche sur les Atomes Froids (IFRAF). Z.~W.
acknowledges support from the European Marie Curie Grant
MIF1-CT-2004-509423, and G.~R. from the DGA.


\end{document}